\documentclass[aps,prx,twocolumn,showpacs,superscriptaddress]{revtex4-1}
\usepackage[english]{babel}
\usepackage{amssymb}
\usepackage{amsmath}
\usepackage{txfonts}
\usepackage{mathdots}
\usepackage[english]{babel} %%% 'french', 'german', 'spanish', 'danish', etc.
\usepackage{amssymb}
\usepackage{amsmath}
\usepackage{txfonts}
\usepackage{mathdots}
\usepackage[classicReIm]{kpfonts}
\usepackage[dvips]{graphicx} %%% use 'pdftex' instead of 'dvips' for PDF output
\usepackage{epsfig}

\begin{document}

\title{Strong magneto-optical response enabled by quantum two level systems}

\author{Lei Ying}
\affiliation{Department of Electrical and Computer Engineering, University of Wisconsin, Madison, Wisconsin 53706, USA}
\author{Ming Zhou}
\affiliation{Department of Electrical and Computer Engineering, University of Wisconsin, Madison, Wisconsin 53706, USA}
\author{Xiaoguang Luo}
\affiliation{Department of Physics, Southeast University, Nanjing 211189, China}
\author{Jingfeng Liu}
\affiliation{College of Electronic Engineering, South China Agricultural University, Guangzhou 510642, China}
\author{Zongfu Yu}
\email{Corresponding author: zyu54@wisc.edu}
\affiliation{Department of Electrical and Computer Engineering, University of Wisconsin, Madison, Wisconsin 53706, USA}

% To be edited by editor
% \dates{Compiled \today}

%\ociscodes{magneto-optical effect; time-reversal breaking; nonreciprocal optics; topological photonics}

% To be edited by editor
% \doi{\url{http://dx.doi.org/10.1364/optica.XX.XXXXXX}}

\begin{abstract}
The magneto-optical effect breaks time-reversal symmetry, a unique property that makes it indispensable in nonreciprocal optics and topological photonics. Unfortunately, all natural materials have a rather weak magneto-optical response in the optical frequency range, posing a significant challenge to the practical application of many emerging device concepts. Here we theoretically propose a composite material system that exhibits an \textit{intrinsic} magneto-optical response orders of magnitude stronger than most magneto-optical materials used today. This is achieved by tailoring the resonant interplay between the quantum electrodynamics of electronic transitions in two-level systems and the classical electromagnetic response of local plasmon resonance.
\end{abstract}

%\setboolean{displaycopyright}{true}

\maketitle

\section{Introduction}

Breaking Lorentz reciprocity using the magneto-optical (MO) effect opens the door to a new class of functionality unattainable in reciprocal systems. It directly contributes to the rise of fields of nonreciprocal \cite{YWF:2007,YF:2009,KMSK:2010,BHJK:2011,LYFL:2012, SCA:2013,ESSA:2014,RMAV:2016} and topological photonics \cite{WCJS:2008,KMTK:2013,LS:2014,ZHLS:2014,SLIY:2015,LFFJ:2016,GLYL:2015,YGTB:2017,LFJS:2013,BNVE:2017}. While many exciting optical sciences are being discovered, such as optical Weyl points \cite{LFJS:2013,NHLC:2017}, quantum Hall effect \cite{HR:2008,RH:2008,HDLT:2011,HMSJ:2013,BSN:2015,SLIY:2015,PBSM:2015,CJNM:2016,ZHGW:2018}, etc., there has always been a challenge: nature does not offer any material with a strong magneto-optical effect in the optical frequency range~\cite{WPRB:1975}. In nonreciprocal photonics, the weak MO effect has been the roadblock in miniaturization of optical isolators, holding back the development of fully integrated Si photonics. In emerging topological photonics, demonstration of nontrivial topological states has rarely gone beyond radio frequency~\cite{WCJS:2009,SLIY:2015}, again because the MO effect of natural materials tails off when transitioning from the radio to the optical frequency. The weak MO response in natural materials prompted the search for other ways of breaking reciprocity, including for example nonlinearity~\cite{GSDM:2009,RKEC:2010,FAHX:2011,FWVS:2012, RZPL:2013,POLM:2014} and time-dependent modulation~\cite{YF:2009,FYF:2012,SCA:2013,ESSA:2014,RMAV:2016}. However, these methods come with their own limitations. The performance of nonlinear methods strongly depends on the intensity of light signal and suffers from dynamical reciprocity~\cite{SYF:2015}. The time-dependent modulation, on the other hand, requires active driving that is both complex and power hungry~\cite{LYFL:2012}.

The lack of strong MO materials has also motivated extensive research on using nanostructures to enhance the MO effect. These structures slow down light to increase interaction time with the MO materials. It is achieved by a variety of resonant structures~\cite{BDZ:2007,GGAN:2008,BOSP:2015,KHJH:2015,MBPS:2016,MYV:2011,AGGG:2008,BAPK:2011,CSWD:2013,KBCN:2013,LMPD:2014,VOPY:2015,LZLQ:2016,ACCG:2015,CKYK:2018,LSCG:2017,MKAA:2014}, including ring resonators, photonic crystal slabs and optical nanoresonators. However, these wavelength-scale structural features substantially limit its general applicability in complex geometries. For example, it is unclear how a ring resonator can be made compatible with three-dimensional Weyl photonic crystals~\cite{LFJS:2013,LWYR:2015}, which come with their own structural specifications.

\begin{figure*}[htbp]
\centering
\epsfig{figure=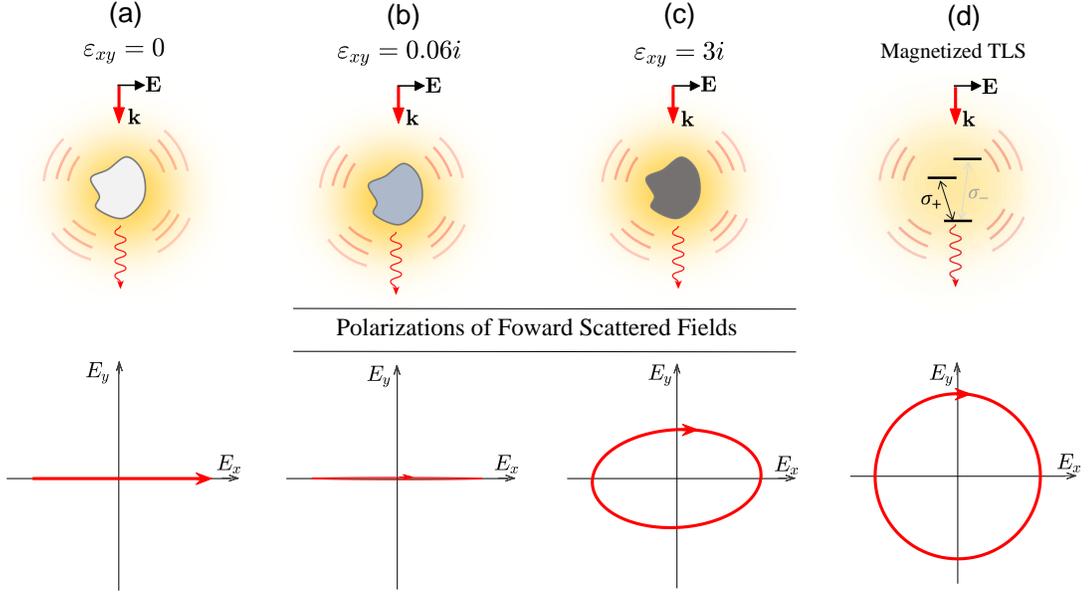,width=6in}
\caption{Schematics of scattered fields and polarizations of forward scattered fields for MO materials with diagonal element ${\varepsilon }_{xx}=6.25$  and off-diagonal dielectric constants (a) ${\varepsilon }_{xy}=0$, (b) ${\varepsilon }_{xy}=0.06i$, and (c) ${\varepsilon }_{xy}=3i$, and (d) a magnetized TLS.  We note that at the deep sub-wavelength scale, the shape and size of the object do not have a significant impact on the results shown above (see confirmation in Sec. I A of supplementary information}
\label{fig:1}
\end{figure*}

An ideal enhancement mechanism should use \textit{intrinsic} material properties instead of \textit{extrinsic} structures that could impede general usability. Here, we theoretically show that a quantum-classical composite material could exhibit an intrinsically large MO response in the optical frequencies. It uses quantum electronic transition in two level systems (TLSs)—such as atoms, molecules, and quantum dots—to provide the intrinsic magneto-optical response. Under an external magnetic field, Zeeman splitting lifts the degenerate optical transitions for left and right circularly polarized light. When the split is greater than the spectral width of individual transitions, left- and right- circularly polarized transitions no longer spectrally overlap, leading to a perfect MO response. This effect can be readily achieved in isolated quantum systems such as TLSs, but not in bulk materials, where a large number of states are densely packed in the energy spectrum and overlapping transitions cancel MO response despite Zeeman splitting. Based on magnetized TLS, we further explore ultra-compact local plasmonic resonance to broaden the bandwidth of enhanced MO effect.  With a strong intrinsic MO response, the material offers unprecedented flexibility in breaking the Lorentz nonreciprocity in the optical frequency range.

\section{Main Results}

The MO strength of a material can be described by the imaginary off-diagonal component of polarizability. Even for some of the strongest MO materials such as yttrium iron garnet (YIG), the magneto-optical response, i.e. the off-diagonal component of the polarizability, is at least three orders of magnitude weaker than the non-MO electromagnetic response, i.e. the diagonal components.\textbf{ }On the other hand, Zeeman splitting of electronic transition is well known to induce a strong magneto-optical effect~\cite{LMK:2001,SBS:2011,PKPW:2016}. When a magnetic field splits the excited state of a TLS, the polarizability tensor at the transition frequency $\sigma_+$ (details in Sec. III C of supplementary information)
\begin{eqnarray}
{\mathop{\boldsymbol{\mathrm{\alphaup}}}\limits^{\leftrightarrow}}_{\mathrm{TLS}}^{(0)}=\frac{\alpha_0}{\sqrt{2}}\ \ \left( \begin{array}{ccc}
1 & i & 0 \\
-i & 1 & 0 \\
0 & 0 & 0 \end{array}
\right),
\end{eqnarray}
where $\alpha_0=3i\varepsilon_0\lambda^3/4\pi^2$. $\varepsilon_0$ is vacuum permittivity and $\lambda$ is the wavelength of the resonant transition. The polarizability exhibits a perfect MO response: the imaginary off-diagonal component has the same magnitude as the diagonal component.

Figure 1 compares MO materials and a magnetized TLS. Unlike bulk materials, where we can use the Faraday rotation angle to measure the MO strength, here we instead use the polarization of the forward scattered light to characterize the strength of the MO response for a single subwavelength object. Specifically, we consider an incident wave linearly polarized along the $x$ direction. When the material has no MO effect (Fig. 1a), the induced polarization $\boldsymbol{\mathrm{p}}$ is linearly polarized along the $x$ direction, resulting in linearly polarized scattered light in the forward direction. For an MO material, e.g. YIG with ${\varepsilon }_{xy}=0.06i$, the induced polarization $\boldsymbol{\mathrm{p}}$ acquires a small component in the $y$ direction with a phase difference of $\pi/2$ relative to that in the \textit{x} direction. This small $y$ component leads to elliptically polarized light as shown in Fig. 1b. Because of the weak MO strength, the polarization is highly elongated along the $x$ direction. If YIG's ${\varepsilon }_{xy}$ were to increase to $3i$, the scattered light would become more circular, as shown in Fig. 1c. These conclusions are independent of the size of the particles, as long as they are well below the wavelength. The strong MO effect of a magnetized TLS can be seen in Fig. 1d, where ${\sigma}^\pm$ represent transitions coupled to clockwise and anti-clockwise circular polarized light, respectively. The scattered fields are circularly polarized, indicating a perfect magneto-optical response.

However, TLS's strong intrinsic MO response suffers from one critical issue: the incident frequency must be around one of the transitions ${\sigma}^\pm$, as shown by the inset of Fig. 1d. Unfortunately, these transitions typically have a very narrow spectral bandwidth ${\gamma }_0$. For example, the bandwidth of the D1 transition (5${}^{2}$S${}_{1/2}$ $\mathrm{\to }\mathrm{\ }$5$^{2}$P${}_{1/2}$) of an Rb87 atom is only $36.1$ MHz~\cite{Steck:2001}. To make this strong MO response relevant for optical applications, the bandwidth must be increased by orders of magnitude.

\begin{figure}[htbp]
\centering
\epsfig{figure=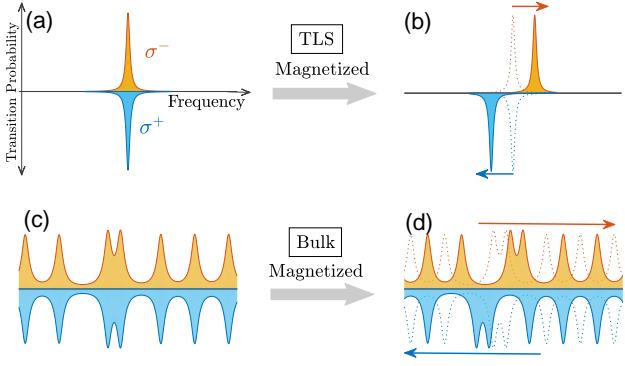,width=\linewidth}
\caption{Schematics of $\sigma^\pm$ transition probabilities for a TLS (a) with and (b) without a magnetic field. (c) and (d) correspond to the bulk system with dense packed $\sigma^\pm$ transitions with same magnetized state to (a) and (b).  When energy levels overlap as shown in (d), the MO response is weak despite the magnetic field.}
\label{fig:2}
\end{figure}

The fundamental limit of bandwidth comes from the small size of the radiating dipoles in TLSs. The relationship between bandwidth and dipole size is given by ${\gamma }_0\sim \mu^2$~\cite{Jackson:1999}. For the D1 transition described above, the effective size of the transition dipole momentum is only 0.08 nm, a rather small size compared to most classical radiators at optical frequencies. Unfortunately, it is rare for a TLS to have a large dipole, due to their small physical sizes. One cannot solve this issue by using super-large TLSs or even bulk materials. The magnetic field splits otherwise degenerate states. As shown in Fig. 2 a, it is essential to have these non-overlapping transitions for clock-wise (orange) and counter-clock-wise (blue) polarized light in order to create MO response. In large systems or bulk materials, the transitions are often densely packed in the energy spectrum. Although the magnetic field can still split the degenerate states, it cannot easily create non-overlapping transition for clock-wise and counter-clock-wise polarized light, as illustrated in Fig. 2b. These overlapping transitions make it difficult to realize strong MO response in large systems or bulk materials~\cite{Pershan:1967} (see Sec. II in supplementary information for further explanation).

\begin{figure}[htbp]
\centering
\epsfig{figure=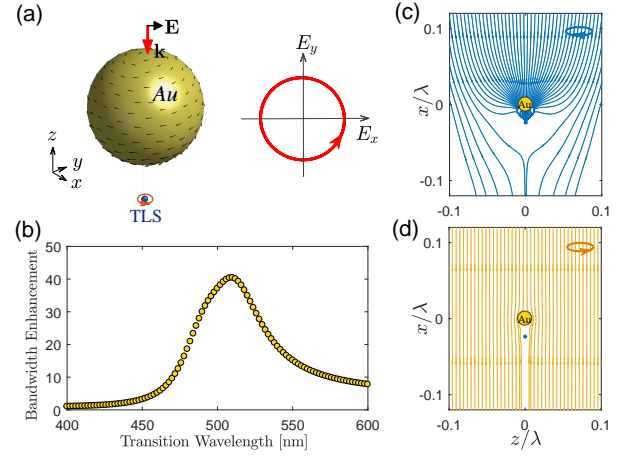,width=\linewidth}
\caption{(a) A magnetized TLS (blue dot) placed below a non-magnetic gold nanoparticle of $5$ nm radius (details in main text). The distance between them is $8$ nm. Black arrows denote the local polarization currents. The red circle shows the polarization of the forward scattering field under a linearly polarized incident wave. (b) The effective broadening of the bandwidth of the $\sigma^+$ transition as a function of transition wavelength. The bandwidth enhancement is defined by $\mathit{\Gamma}/{\gamma}_0$. (c) and (d) The Poynting flux lines around the magnetized TLS-nanoparticle (gold) composite. The frequency of the incident light is tuned to be the frequency of transition $\sigma_+$, i.e. $\omega=\omega_0-\Delta\omega_{\mathrm B}/2$. Great contrast can be seen in the optical cross section when under different polarization, even at the same frequency. All calculations are based on exact method in Eqs. (2) and (3).   }
\label{fig:3}
\end{figure}

In classical electromagnetics, the dipole moment may not always be limited by small physical size. For example, considering a spherical nanoparticle with a subwavelength size (e.g. 5 nm), the induced dipole moment is~\cite{KCZS:2003,CRFD:2998} $\boldsymbol{\mathrm{p}}\approx 3V\frac{{\varepsilon }_{\mathrm{NP}}-1}{{\varepsilon }_{\mathrm{NP}}+2}{\varepsilon }_0{\boldsymbol{\mathrm{E}}}_0$, where \textit{V} and ${\varepsilon }_{\mathrm{NP}}$ are the volume and the permittivity of the nanoparticle, respectively, and $\varepsilon_0$ is the vacuum permittivity. The dipole moment diverges when ${\varepsilon }_{\mathrm{NP}}=-2$, resulting in an effective dipole of infinite size. This effect contributes to local plasmonic resonances that are widely used for sensing~\cite{NJY:1996,OYK:2000}. In practice, the dipole moment will not diverge due to the loss in real materials. Using this effect, we show a quantum-classical composite system~\cite{ZGB:2006,TH:2008,Govorov:2010,KPLL:2012,VKH:2012,CSA:2013,GHMT:2013,DFBG:2014,YPL:2015} that can exhibit a magnetized dipole of greatly enhanced size. Figure 3a shows one example of such a composite system: it consists of a magnetized TLS placed near a non-magnetic nanoparticle. The distance between the TLS and the nanoparticle is relatively short ($<$ 20 nm).

First, we provide an intuitive understanding of how the composite system helps to increase the effective size of the dipole in a magnetized TLS, and consequently, its bandwidth. We again consider the forward scattering field under a linearly polarized incident wave. The magnetized TLS produces a very strong scattered field around itself. The field as felt by a nearby nanoparticle is thus dominated by this scattered field, which is circularly polarized. As a result, the induced polarization in the nanoparticle is circular, instead of following the incident field to create a linear polarization. Figure 3a illustrates the induced polarization current of the nanoparticle. This circulating current produces a circularly polarized scattered field, which further contributes to the MO response of the system. It effectively increases the size of the rotating dipole of the magnetized TLS, leading to a much stronger coupling to the radiation continuum, and thus a broader radiative bandwidth, than a bare TLS alone.

Another perspective to understand bandwidth broadening is the Purcell factor calculated as~\cite{GMLF:2005,CSA:2013} $\mathcal{F}=\frac{\mathit{\Gamma}}{{\gamma }_0} = \frac{{\boldsymbol{\mathrm{\muup }}}^{\dagger}\mathrm{\cdot }\mathrm{Im}\mathop{\boldsymbol{\mathrm{G}}}\limits^{\leftrightarrow}\left({\boldsymbol{\mathrm{r}}}_{\mathrm{TLS}},{\boldsymbol{\mathrm{r}}}_{\mathrm{TLS}}\right)\cdot\boldsymbol{\mathrm{\muup }}}{{\boldsymbol{\mathrm{\muup}}}^{\dagger}\cdot\mathrm{Im}{\mathop{\boldsymbol{\mathrm{G}}}\limits^{\leftrightarrow}}_{\boldsymbol{0}}\left({\boldsymbol{\mathrm{r}}}_{\mathrm{TLS}},{\boldsymbol{\mathrm{r}}}_{\mathrm{TLS}}\right)\mathrm{\cdot }\boldsymbol{\mathrm{\muup }}}\boldsymbol{\ }$, where $\mathop{\boldsymbol{\mathrm{G}}}\limits^{\leftrightarrow}={\mathop{\boldsymbol{\mathrm{G}}}\limits^{\leftrightarrow}}_0+{\mathop{\boldsymbol{\mathrm{G}}}\limits^{\leftrightarrow}}_s$ is the dyadic Green functions~\cite{Tai:1994}. ${\mathop{\boldsymbol{\mathrm{G}}}\limits^{\leftrightarrow}}_{0,s}$ are the Green's function in free space and the scattered Green's function due to the nanoparticle, respectively. The bandwidth enhancement diverges when the permittivity ${\varepsilon }_{\mathrm{NP}}\sim -2$.  For real materials, such as gold, the optical loss curtails the divergence but still produces significant bandwidth enhancement. An example of a gold nanoparticle is shown in Figs. 3a\&b. From the Purcell factor alone, one cannot conclude if the composite system's response is dominated by the TLS's MO response or the non-magnetic response of the nanoparticle. The modified decay rate $\mathit{\Gamma}$ includes both radiative and non-radiative parts.

Next, we perform quantum electrodynamic modeling of the composite system to explicitly show the MO response. When a plane wave is incident upon a composite TLS-nanoparticle system, the expectation value of the total field operator $\widehat{\boldsymbol{\mathrm{E}}}\left(\boldsymbol{\mathrm{r}}\right)$ can be written as~\cite{YPL:2015}
\begin{equation}
\left\langle \widehat{\boldsymbol{\mathrm{E}}}\left(\boldsymbol{\mathrm{r}}\right)\right\rangle ={\boldsymbol{\mathrm{E}}}_0\left(\boldsymbol{\mathrm{r}}\right)+{\boldsymbol{\mathrm{E}}}^s_0\left(\boldsymbol{\mathrm{r}}\right)+\left\langle {\widehat{\boldsymbol{\mathrm{E}}}}_{\mathrm{TLS}}\left(\boldsymbol{\mathrm{r}}\right)\right\rangle \mathrm{,}
\end{equation}
where ${\boldsymbol{\mathrm{E}}}_0$ represents the incident plane wave and ${\boldsymbol{\mathrm{E}}}^s_0$ is the scattered field of the nanoparticle in the absence of the TLS. The scattered field operator of the TLS is given by $\left\langle {\widehat{\boldsymbol{\mathrm{E}}}}_{\mathrm{TLS}}\left(\boldsymbol{\mathrm{r}}\right)\right\rangle \mathrm{=}\left({\omega }^2/{\varepsilon }_0c^2\right)\mathop{\boldsymbol{\mathrm{G}}}\limits^{\leftrightarrow}\left(\boldsymbol{\mathrm{r}},{\boldsymbol{\mathrm{r}}}_{\mathrm{TLS}},\omega \right)\mathrm{\cdot }\boldsymbol{\mathrm{d}}$, where the induced dipole moment \textbf{d} can be solved by a Bloch equation under weak excitation~\cite{CSA:2013,YPL:2015} (also see details in Sec. III A of supplementary information)
\begin{equation}
\boldsymbol{\mathrm{d}}\boldsymbol{\mathrm{(}}\omega \boldsymbol{\mathrm{)}} = \frac{\mathrm{1}}{\mathrm{\hbar }}\sum_m{\frac{{\boldsymbol{\mathrm{\muup }}}^{\dagger}_m\mathrm{\cdot }{(\boldsymbol{\mathrm{E}}}_0+\boldsymbol{\mathrm{E}}_0^s)}{\omega -{\omega}_m+\delta {\omega }_m\mathrm{+}i\mathit{\Gamma}_m/2}{\boldsymbol{\mathrm{\muup}}}_m}={\mathop{\boldsymbol{\mathrm{\alphaup}}}\limits^{\leftrightarrow}}_{\mathrm{TLS}}^{\mathrm{\left(eff\right)}}\cdot \boldsymbol{\mathrm{E}}_0,
\end{equation}
where $m=\pm$ denote the transitions $\sigma^\pm$ and $\omega$ is the frequency of the incident wave. The effective TLS polarizability ${\mathop{\boldsymbol{\mathrm{\alphaup}}}\limits^{\leftrightarrow}}_{\mathrm{TLS}}^{\mathrm{\left(eff\right)}}$ has been dressed by nanoparticle (see Sec. III D in supplementary information).
The transition frequency ${\omega}_m+\delta {\omega}_m$ includes the Lamb shift $\delta \omega_m =\left({\omega}^2/{\varepsilon}_0{\hbar}^2\right){\boldsymbol{\mathrm{\muup }}}^\dagger_m\mathrm{\cdot }$Re$\mathop{\boldsymbol{\mathrm{K}}}\limits^{\leftrightarrow}\left({\boldsymbol{\mathrm{r}}}_{\mathrm{TLS}},{\boldsymbol{\mathrm{r}}}_{\mathrm{TLS}},\omega \right)\mathrm{\cdot }{\boldsymbol{\mathrm{\muup }}}_m$. The classical electromagnetic response of the nanoparticle is included in the Green's function ~\cite{TH:2008,GHMT:2013}$\mathop{\boldsymbol{\mathrm{K}}}\limits^{\leftrightarrow}\left(\boldsymbol{\mathrm{r}},{\boldsymbol{\mathrm{r}}}^{\mathrm{'}},\omega \right)\mathrm{=}\mathop{\boldsymbol{\mathrm{G}}}\limits^{\leftrightarrow}\left(\boldsymbol{\mathrm{r}},{\boldsymbol{\mathrm{r}}}^{\mathrm{'}},\omega \right)\mathrm{-}\left(c^2/{\varepsilon }_0{\omega }^{\mathrm{2}}\right)\delta \left(\boldsymbol{\mathrm{r}}\mathrm{-}{\boldsymbol{\mathrm{r}}}^{\mathrm{'}}\right)\mathop{\boldsymbol{\mathrm{I}}}\limits^{\leftrightarrow}$.

Under a linearly polarized incident wave, we calculate the polarization of the scattered field in the forward direction for the cases studied in Fig. 3a. The polarization is almost circular, indicating strong MO response. The induced current calculated on the surface of the nanoparticle is indeed circulating, as shown in Fig. 3a.
The circular polarization of a TLS near a nanoparticle can be described by the degree of circular polarization (DOCP), which levels off to $+100$\% at transition frequency $\sigma^+$ (see details in Sec. III D of supplementary information). It indicates the polarization of this modified TLS is perfectly circular.
The strong MO response can also be seen under circularly polarized incidence. At the $\sigma^+$transition frequency, the clock-wise polarization strongly excites the composite, while the other, orthogonal polarization barely interacts with the composite (as shown as the Poynting flux flows in Figs. 3c\&d).

Next, we consider a TLS with a specific radiative linewidth ${\gamma }_0=1\ \mathrm{GHz}$~\cite{SJKH:2009} around $550$ nm wavelength. To allow some spacing between the TLS and the gold particle, we embed the TLS in a uniform background material with a dielectric constant ${\varepsilon }_b=2.25$. The permittivity of  gold nanoparticle ${\varepsilon }_{NP}\ $ is given by the Drude model with non-local effect and Landau damping (Sec. III F in supplementary information). In general, as the TLS-nanoparticle spacing decreases, the bandwidth broadens while the peak value of the off-diagonal polarizability decreases. The spacing is 8 nm and the radius of the gold particle is 5 nm. In practice, such structures can be produced using core-shell structures~\cite{OAWH:1998,LSC:2003,LE:2006}. We apply a finite magnetic field corresponding to a Zeeman splitting of $\mathit{\Delta}{\omega }_{\mathrm B}=200{\gamma}_0$. Here we use full-wave electromagnetic modeling in Eqs. (2) and (3). The Mie scattering field $\boldsymbol{\mathrm{E}}_0^s$ and the dyadic Green's function $\mathop{\boldsymbol{\mathrm{G}}}\limits^{\leftrightarrow}\left(\boldsymbol{\mathrm{r}},{\boldsymbol{\mathrm{r}}}^{\mathrm{'}},\omega \right)$ can be numerically calculated~\cite{FS:2001,Tai:1994}. Figure 4a shows the off-diagonal component of the polarizability ${\alpha }_{xy}$ with and without the nanoparticle. Two peaks correspond to the two Zeeman-splitting transitions. The ratio $|{\alpha }_{xy}|/\left|{\alpha }_{xx}\right|\ \ $ remains close to 1 at resonant frequencies of transitions ${\sigma }^\pm$ because the bandwidth broadening is less than the Zeeman splitting and there is almost no overlap between these two transitions. The nanoparticle broadens the bandwidth by 40 times, and the bandwidth broadening also varies with the direction of incident light. Here we obtain greater enhancement in Fig. 4a than that in Fig. 3b because the incident direction is normal to the axis that connects the TLS and the center of the nanoparticle.

The enhanced bandwidth of 40 GHz is quite useful for communication applications. It would be important to further enhance the bandwidth to the THz range to enable broader applications. This cannot be achieved by reducing the spacing between the TLS and the nanoparticle, as the MO response will be weakened by the strong ohmic loss in metal.

\begin{figure*}[htbp]
\centering
\epsfig{figure=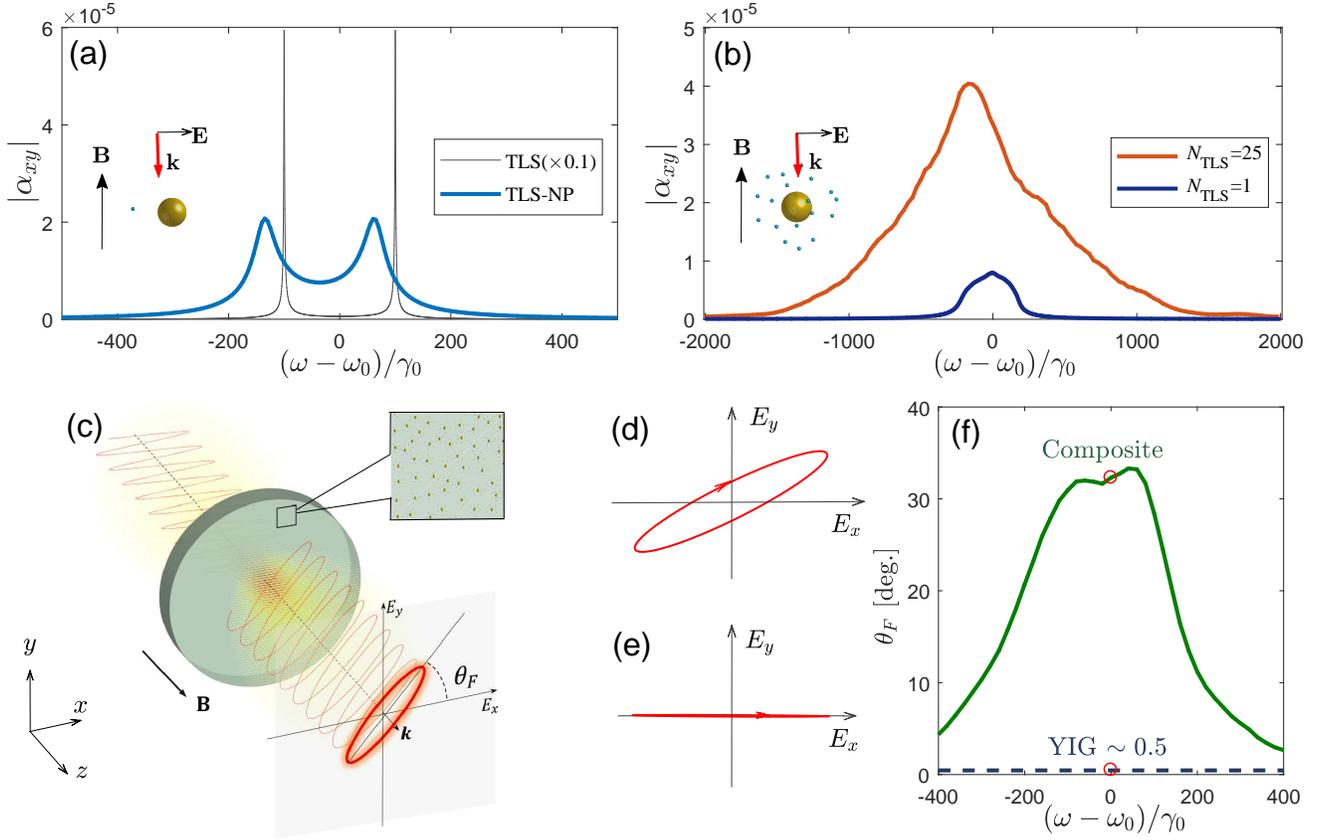,width=\linewidth}
\caption{(a) The  off-diagonal component of the polarizability \textit{$\left|{\alpha }_{xy}\right|$} with (blue) and without (grey) nanoparticle (NP). A TLS is at left side of nanoparticle with same distance and radius to Fig. 3a. (b) Inset shows a quantum-classical composite system with multiple TLSs. The blue-circle and red curves correspond to the off-diagonal component of polarizability \textit{$\left|{\alpha }_{xy}\right|$} for $1$ TLS with frequency distribution over a bandwidth of $\mathit{\Delta}\omega =200{\gamma }_0$ and $25$ TLSs with \textit{$\mathit{\Delta}\omega =1250{\gamma }_0$}, respectively. Due to randomization and averaging, the split peaks as shown in (a) disappear. (c) Faraday rotation of linearly polarized light passing through a disk filled with magnetized TLS-nanoparticle composites. Inset shows the zoom-in view of the glass material filled with composites. (d) and (e) The polarizations of outgoing fields for the composite and conventional YIG (\textit{${\varepsilon}_{xx}=6.25$} and \textit{${\varepsilon}_{xy}=0.06i$} ) of the same geometry, respectively. (f) The Faraday rotation angles of the composite and YIG as a function of the frequency. Upper and lower red circles correspond to (d) and (e), respectively. The Zeeman splitting energy is fixed at $200\gamma_0$.}
\end{figure*}

Next, we use multiple TLSs to further broaden the bandwidth by at least one order of magnitude. Specifically, a cluster of TLSs is randomly placed on a sphere centered around the nanoparticle, as shown by the inset of Fig. 4b. The transition frequencies of these TLSs are not necessarily to be identical, and are randomly distributed in a spectral range $\mathrm{\Delta }\omega $. Because of this frequency distribution, multiple transitions at different frequencies enable further enhancement of the MO bandwidth. As a result, it can be greater than the Zeeman splitting. We perform quantum electrodynamic modeling that fully incorporates the electrodynamic interactions among all components in the composite system with the radiative correction to the dipole approximation for nanoparticles; the method is shown in Sec. III F of supplementary information. We compare the MO response for the cases of one and $25$ TLSs, which are shown by the blue and red curves in Fig. 4b, respectively. The spectral bandwidth for ensemble case is defined as the area of the off-diagonal polarizability spectrum divided by its maximum value ${\int |\alpha_{xy}(\omega)|d\omega}/{{\max}|\alpha_{xy}(\omega)|}$ and it reaches 1300${\gamma }_0$ for the 25 TLSs case, which is equivalent to $1.3$ THz for TLSs. Over this broadened bandwidth, the ratio between the diagonal and off-diagonal polarizability remains high at around 1/2. The polarizability tensor of multiple TLSs is evaluated by $\mathop{\boldsymbol{\mathrm{\alphaup }}}\limits^{\leftrightarrow}\approx \sum^{N_{\mathrm{TLS}}}_{i=1}{{\mathop{\boldsymbol{\mathrm{\alphaup }}}\limits^{\leftrightarrow}}^{\left(i\right)}}_{\mathrm{TLS}}$, where $N_{\mathrm{TLS}}$ is the number of TLSs. ${\mathop{\boldsymbol{\mathrm{\alphaup }}}\limits^{\leftrightarrow}}^{\left(i\right)}_{\mathrm{TLS}}$ is the dressed polarizability of the $i$th TLS and it is solved by considering all radiative interactions(see details in Sec. III F of supplementary information). Such a direct sum is a reasonable estimate due to the deep subwavelength scale of the composite. Given the randomness of the location and transition frequency of the TLSs, we average the result for 1000 random simulations, each having a different random configuration. We also note that without the nanoparticle a cluster of magnetized TLSs alone would still exhibit extremely narrow bandwidths.

The TLS-nanoparticle composite shown here has a typical radius of around 10 nm, which is small enough that a large number of composite units can be incorporated into a bulk material. It can then be fabricated into a complex structure with little geometrical constraint. So far, we have only analyzed a single composite. To support the above claim, we now provide an example of such a bulk material with greatly enhanced MO response.

We consider a glass material with a refractive index of 1.5. To enable a strong MO response in this glass, we embed magnetized TLS-nanoparticle composites. Each composite has a $5$ nm gold nanoparticle and $\mathrm{10}$ TLSs that are 8 nm away from the nanoparticle surface. We perform full electrodynamic simulation of the material. This is a computationally expensive simulation, considering that we model the response of each TLS and nanoparticle without using any effective medium assumption. Accordingly, we must limit the size of the material to make the computation feasible.

Here, we consider a disk with a radius of 550 nm and height of 45 nm, filled with about 300 composites randomly distributed (Fig. 4c). In practice, we can use more composites and more TLSs to further increase the MO strength. We consider the Faraday rotation of a Gaussian beam passing through the material (details of modeling shown in Sec. V of supplementary
information). The beam wrist is half of the wavelength, so the beam can maximally fit onto the disk. The Zeeman splitting caused by external magnetic field is $\mathrm{\Delta }{\omega }_{\mathrm B}\mathrm{=200}{\mathrm{\gammaup}}_0$, as before. Light transmitted through the disk is generally elliptically polarized. The Faraday rotation angle is estimated as $\theta_F = (1/2) \mathrm{atan} \left[ 2\mathrm{Re} \left( E_y/E_x \right) /\left( 1-|E_y/E_x|^2\right)  \right]$ are the electrical fields of the transmitted light in the \textit{x} and \textit{y} directions, respectively. Despite the ultra-thin thickness of the disk, the Faraday rotation reaches 30 degrees, as shown in Fig. 4b. In contrast, for a YIG disk of the same size, the Faraday rotation angle is 0.5 degrees (Fig. 4d-f). The Faraday rotation of the glass remains almost two orders of magnitude greater than that of the YIG material for a spectral bandwidth of 0.1 nm. Here the Faraday rotation through the YIG disk is performed by using finite-element methods to solve the Maxwell's equations.

\section{Discussions}

Magnetized TLS-nanoparticles can realize an \textit{intrinsic} MO response that is two orders of magnitude stronger than that of natural materials such as YIG. In contrast, all existing enhancement methods do not modify any fundamental property of a material. Their reliance on extrinsic structure features substantially limits general usability. The magnetized TLS-nanoparticle is also within the reach of experimental realization. For example, CdSe/CdTe quantum dots can be easily fabricated~\cite{NSMB:1994,NB:1996} and their transition frequency tuned by size.
The Zeeman splitting energy of semiconductor quantum dots caused by the spin splitting has been expressed as $\Delta\omega_{\mathrm B}=g\mu_BB/\hbar$~\cite{KBWK:1999}, where $g$ is the Landé g-factor. The Zeeman splitting in our calculation $\Delta \omega_{\mathrm B}=200\gamma_0$ corresponds to an external magnetic field $\mathrm{B=0.56T}$.
In practice, many TLSs have a cross section well below $3\lambda^2/2\pi$. We show in Sec. III E of supplementary information that the current on the nanoparticle can still be circularly polarized.
In addition, it is quite feasible to fabricate nanoparticles that are attached with quantum dots. Examples can be seen in Ref.~\cite{KPLL:2012}.

Due to computation power limitations, we only show an example with operation at around 550 nm wavelength and a bandwidth of 0.1 nm. However, in practice, both the operation wavelength and bandwidth can be greatly extended. The operation wavelength is determined by the local plasmon resonance of the spherical gold particle. By tuning core shell structures, the local plasmon resonance can be realized in the full optical (400--800 nm) wavelength range~\cite{OAWH:1998,LSC:2003,LE:2006}.

It is also straightforward to broaden the bandwidth to tens or even hundreds of nanometers. To do so, we can use the inhomogeneous broadening of the TLS-nanoparticle composites: the radii of the nanoparticles and the size of the TLS can be made to distribute over a broad spectral bandwidth. In addition, the spectral bandwidth also scales linearly with the number of TLSs, providing another way to enhance the bandwidth (see details in Sec. IV of supplementary information). The optical loss of the composite material is mainly contributed by the absorption by nanoparticles. The example shown in Fig. 4 has $20\%$ optical absorption with a Faraday rotation angle of 20 degrees. This is remarkably good considering that the disk is only 45 nm thick. The loss is small even compared to the all-dielectric on-chip isolator~\cite{BHJK:2011}.  The loss can also be tuned by the density of the composite in the hosting material. Lastly, we note that the MO response has been reported for a noble metal nanoparticle coated with transition metal \cite{JXWC:2009,WCHC:2011}. Its MO response is two orders of magnitude weaker than the magnetized TLS-nanoparticle shown here (details in Sec. I B of supplementary information).

%\section*{Funding Information}
%National Science Foundation (NSF) (EFRI NewLAW Award 1641109); The DARPA DETECT program.

\section*{Acknowledgments}
The authors acknowledge the financial support of NSF (EFRI NewLAW Award 1641109) and the DARPA DETECT program.

\end{document}